\documentclass[
    ,final            
  ]
  {aipproc}

\layoutstyle{6x9}

\begin{document}

\title{Nonlinear shock acceleration and $\gamma$-ray emission from Tycho and
Kepler}


\author{G. Morlino}{
  address={INAF - Osservatorio Astrofisico di Arcetri, L.go E. Fermi 5, I -
           50125 Firenze, Italia}
}

\author{D. Caprioli}{
  address={Department of Astrophysical Sciences, Princeton University,
           Princeton, NJ 08544, USA}
}

\begin{abstract}
We apply the non-linear diffusive shock acceleration theory in order to
describe the properties of two supernova remnants, SN 1572 (Tycho) and SN
1604 (Kepler). 
By analyzing the multi-wavelength spectra, we infer that both Tycho's and
Kepler's forward shocks (FS) are accelerating protons up to $\sim 500$ TeV,
channeling into cosmic rays more than 10 per cent of their kinetic energy.
We find that the streaming instability induced by cosmic rays is consistent with
the X-ray morphology of the remnants, indicating a very efficient magnetic
field amplification (up to $\sim 300 \mu$G).
In the case of Tycho we explain the $\gamma$-ray spectrum from the GeV up to the
TeV band as due to pion decay produced in nuclear collisions by accelerated
nuclei scattering against the background gas. On the other hand, due to the
larger distance, the $\gamma$-ray emission from Kepler is not detected, being
below the sensitivity of the present detectors, but it should be detectable by
CTA.
\end{abstract}

\maketitle


\section{Introduction}
Modelling particle acceleration at SNRs is one of the main goal to explain the
origin of Galactic cosmic rays (CRs) in the context of the so-called
\emph{supernova paradigm} \cite[see e.g.][]{dav94}. This paradigm requires that
SNRs are able to accelerate nuclei up to energies as high as a few times $10^6$
GeV, converting a fraction $\sim10\%$ of the SNR kinetic energy into CRs.
There is an increasing amount of evidence that shocks in young SNRs can indeed
reach the required efficiency. Unfortunately, all these evidence are indirect
and when considered individually can be also explained by other mechanisms that
do not require efficient acceleration.
A clear example of such an ambiguous situation is represented by $\gamma$-ray
emission. In the last few years several SNRs have been detected both in the
GeV and in the TeV band but, in spite of this increasing amount of data,
the question whether this emission is due to hadronic (through the decay of
neutral pions produced in nuclear interactions between accelerated nuclei and
the background plasma) or leptonic processes (due to inverse Compton and/or
bremsstrahlung) is still debated \citep[see e.g.][for a general discussion on
this topic]{Ellison07}.
A possible way to discriminate between the \emph{leptonic} (i.e. inefficient)
and the \emph{hadronic} (i.e. efficient) scenario is studing the  effects that
particle acceleration produces at all observable wavelengths, rather that focus
the attention only on the $\gamma$-ray emission, simultaneously using multiple
set of data to constrain the model.


Here we apply the non-linear diffusive shock acceleration (NLDSA) theory in
order to explain the non-thermal emission and morphological properties of two
similar SNRs, namely Tycho and Kepler.
Tycho, in particular, has been recently detected in $\gamma$-rays by Fermi-LAT
\cite[]{Giordano11} and VERITAS \cite[]{Acciari11}, and can be considered one of
the most promising object where to test the shock acceleration theory and hence
the CR--SNR connection. We presented a detailed model applyed to Tycho in
\cite{morlino11}. Kepler, on the other hand, has not been detected in
$\gamma$-ray band, yet, but it is very similar to Tycho in many respects, and we
show that the predicted $\gamma$-ray spectrum should be detectable by CTA.

\section{Coupling particle acceleration with remnant evolution}
Both Tycho and Kepler have been established to be remnants of a type Ia SN,
Tycho thanks to observation of the scattered-light echo \cite[]{Krause08}
while Kepler based on the O/Fe ratio observed in the X-ray spectrum
\cite[]{Reynolds07}. They have similar ages and similar values for the radio
spectral index, i.e. 0.65 and 0.64, respectively. Interestingly also the ratio
of power emitted at 10 keV and at 1 GHz, i.e. $\nu F_{\nu}(@10{\rm keV})/\nu
F_{\nu}(@1\rm{GHz})$, is very similar, being 56.9 for Tycho and 52.9 for Kepler.
Finally, both remnants present very thin filaments in non-thermal X-ray
emission.

In order to make the model as simple as possible, in both cases we assume that
the remnants expand into an uniform circumstellar medium (CSM) with proton
number density $n_{0}$ (which we leave as a free parameter) and temperature
$T_{0}=10^{4}~{\rm K}$.
We model the remnant evolution by following the analytic prescriptions given
by \cite{TMK99} for Type Ia SNe, assuming a SN explosion energy $E_{SN}=10^{51}$
erg and one solar mass in the ejecta, whose structure function is taken as
$\propto (v/v_{ej})^{-7}$.
The radial structure of density and temperature profiles is then calculated
by assuming that the shocked CSM is roughly in pressure equilibrium.

On top of this SNR evolution, the spectrum of accelerated particles is
calculated according to the semi-analytic kinetic formalism put forward in
\cite{dsax0} and references therein, which solves self-consistently the
equations for conservation of mass, momentum and energy along with the
diffusion-convection equation describing the transport of non-thermal particles
for quasi-parallel, non-relativistic shocks. The injection is regulated by a
free parameter, $\xi_{\rm inj}$, in such a  way that all particles from the
thermal plasma with momentum $p>\xi_{\rm inj} p_{th,2}$, with $p_{th,2}$ the
typical thermal momentum downstream, start the acceleration process.

A crucial role in our model is played by the magnetic field amplification
induced by the super-Alfv\'enic streaming of relativistic particles upstream of
the shock. We model this magnetic amplification as in \cite{jumpkin}.
The large magnetic fields predicted by the resonant amplification have two
main consequences: 1) the magnetic pressure upstream becomes comparable to, or
even larger than, the thermal plasma pressure and, reducing the
compressibility of the plasma, affects the shock compression factor.
2) When the magnetic field is amplified the velocity of the scattering centers,
which is generally neglected with respect to the shock speed, may be
significantly enhanced \cite[]{jumpkin}. When this occurs, the total
compression factor felt by accelerated particles may be appreciably reduced and,
in turn, the spectra of accelerated particles may be considerably softer.
We explicitly include these effects assuming that the scattering centers moves
with a speed equal to the Alfv\'en speed in the amplified magnetic field.
Once the magnetic field structure is known, we can compute the spectrum of
accelerated electrons at the shock, which is assumed to be proportional to the
proton spectrum $f_e(p)= K_{ep}f_p(p)$, up to a maximum momentum $p_{e,\max}$
determined by the synchrotron losses in the amplified magnetic field, where the
spectrum presents a squared exponential cut-off $\propto
\exp{[-(p/p_{e,\max})^2]}$.
The evolution of the electron and proton spectrum downstream of the shock is
computed taking into account adiabatic losses for protons and adiabatic and
synchrotron losses for electrons.

{\bf RESULTS.}
We consider the following radiative processes: 1) synchrotron emission of
relativistic electrons; 2) thermal and non-thermal electron bremsstrahlung; 3)
inverse Compton scattering (ICS) of electrons on CMB radiation, local IR dust
emission and Galactic IR+optical light; 4) emission due to the decay of $\pi^0$ 
produced in hadronic collisions.
Because the explosion energy and the mass of the ejecta are fixed {\it a
priori}, in order to fit the observed spectra and the remnant angular size we
can only vary three parameters: the number density of the upstream medium,
$n_0$, the injection efficiency of protons, $\xi_{\rm inj}$, and the electron to
proton normalization, $K_{ep}$. 

\begin{figure}
  \includegraphics[height=.2\textheight]{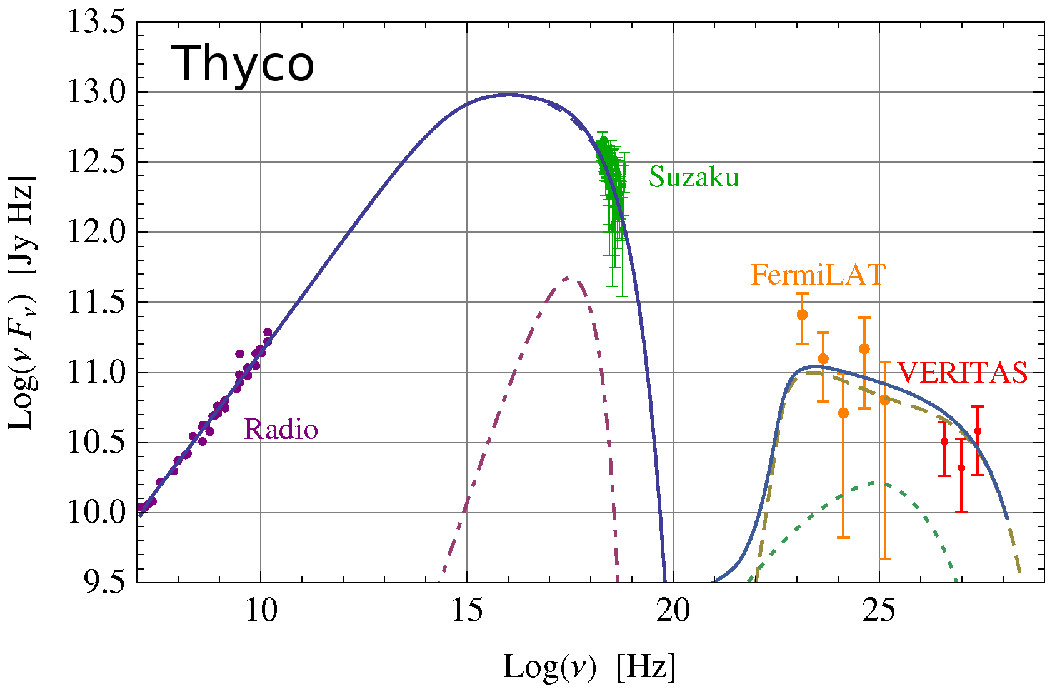}
  \includegraphics[height=.2\textheight]{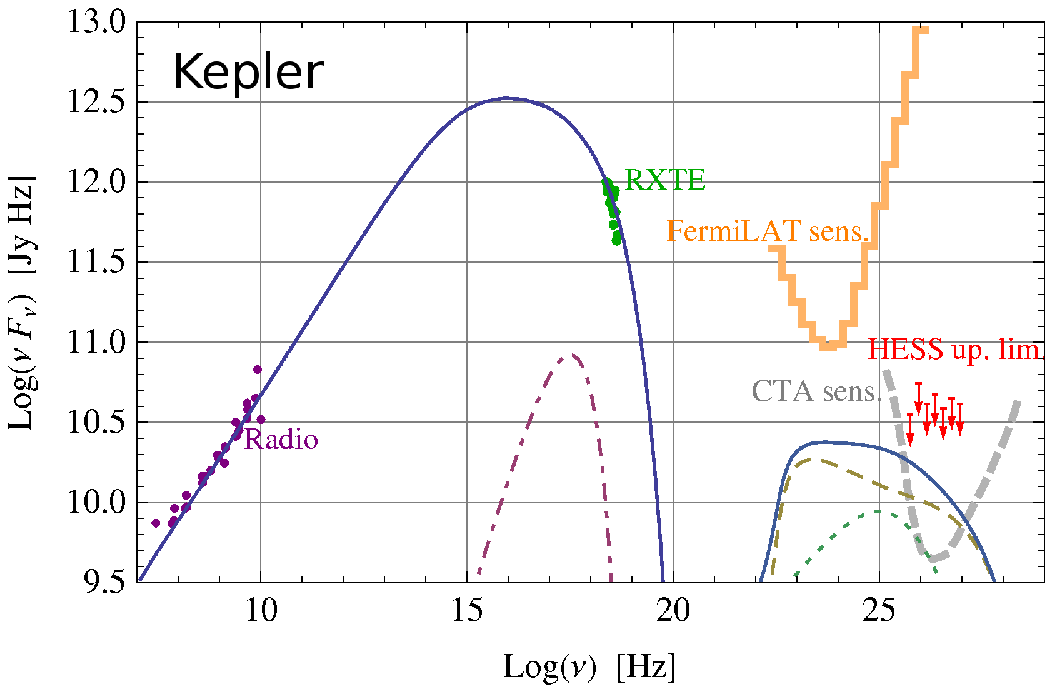}
  \caption{Spatially integrated spectral energy distribution of Tycho (left
   panel) and Kepler (right panel). The curves show synchrotron emission (thin
   dashed), thermal electron bremsstrahlung (dot-dashed), $\pi^0$ decay (thick
   dashed) and ICS (dotted) as calculated within our model. The total emission
   is showed by the solid curve. The experimental data for Tycho are,
   respectively: radio from {\protect\cite{ReyEllison92}}; X-rays from Suzaku
   (courtesy of Toru Tamagawa), GeV $\gamma$-rays from Fermi-LAT
   {\protect\cite[]{Giordano11}} and TeV $\gamma$-rays from VERITAS
   {\protect\cite[]{Acciari11}}. For Kepler: radio from
   {\protect\cite{ReyEllison92}} and X-rays from {\protect\cite{allen99}}.}
   \label{fig:nuF_nu}
\end{figure}

In Fig.~\ref{fig:nuF_nu} we show our best fit of the photon spectrum
produced by the superposition of all the radiative processes outlined above,
comparing it with the existing data. The overall agreement is quite good.
Quite interestingly, for both Tycho and Kepler our best-fitting returns
$n_0=0.3\, \rm{cm^{-3}}$ and $\xi_{\rm inj}= 3.7$ which implies that the total
amount of kinetic energy channeled into CRs is $\simeq 12\%$. The inferred
values for the circumstellar medium implies that Tycho and Kepler have a
distance of 3.3 and 7.4 kpc from the Sun.

\begin{figure}
  \includegraphics[height=.2\textheight]{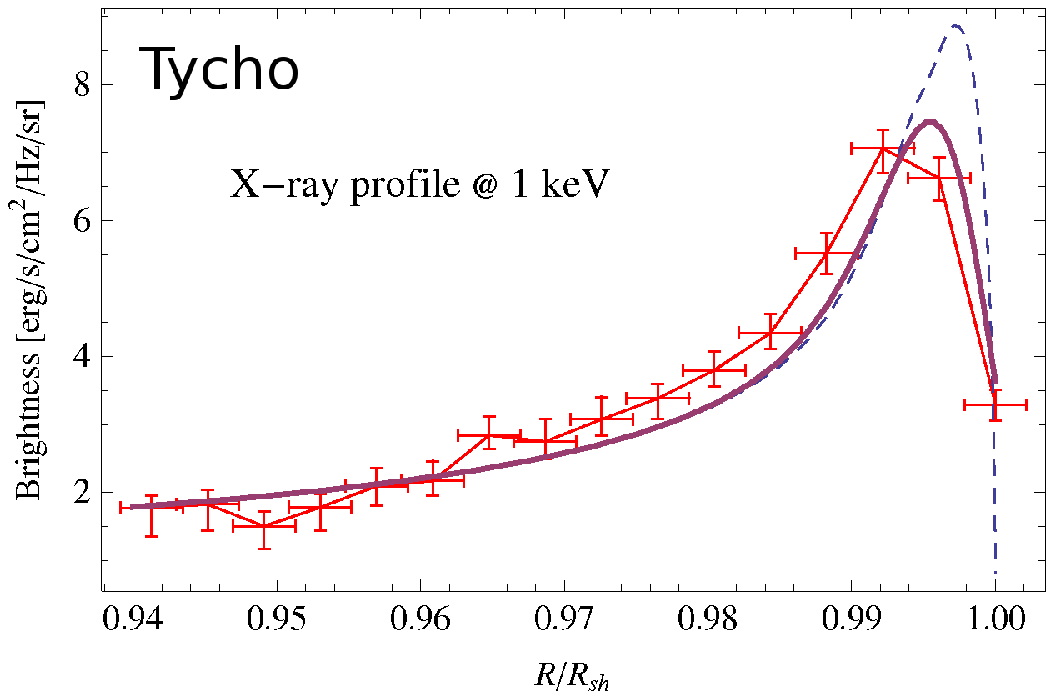}
  \includegraphics[height=.2\textheight]{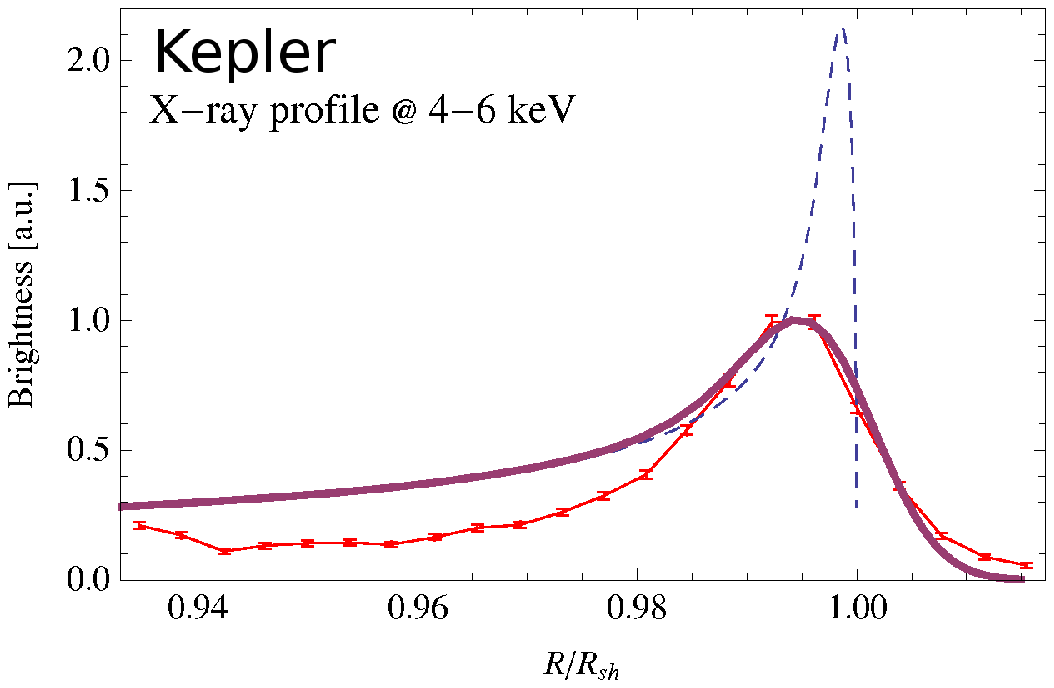}
  \caption{ Projected X-ray emission at 1 keV for Tycho (left panel) and at 4-6
   keV for Kepler (right panel) as a function of the distance from the shock
   position. The {\it Chandra} data points are from {\protect \cite{Gamil07}}
   (for Tycho) and from {\protect \cite{Vink08}} (for Kepler). The dashed lines
   shows the projected radial profile of synchrotron emission while the solid
   lines are the same but convolved with the {\it Chandra} point spread function
   ($\sim0.5''$).
  }
  \label{fig:Xray}
\end{figure}

Synchrotron radiation fits well the observed X-ray emission assuming an electron
to proton ratio $K_{ep}= 1.6\times 10^{-3}$ for Tycho and $2.8\times 10^{-3}$
for Kepler.
The projected X-ray emission profile is shown in Fig.\ref{fig:Xray}, where it is
compared with available Chandra data. In the case of Tycho the predicted profile
shows a  very good agreement with the data while for Kepler the model
overpredicts the emission in the inner part of the rim. This could be due to a
deviation of the shock geometry from the perfectly spherical symmetry assumed
in our model.
The sharp decrease of the emission behind the FS is due to the rapid synchrotron
losses of the electrons in a magnetic field as large as $\sim 300 \mu$G.

In the $\gamma$-ray band the decay of $\pi^0$ produced in hadronic
collision is the dominant process. In particular, we predict a slope for
accelerated protons $\propto E^{-2.2}$. In the case of Tycho this slope well
accounts for Fermi-LAT and VERITAS data within the experimental errors.
The predicted proton spectrum shows a cut-off around $p_{\max} = 470$ TeV/c.

ICS of relativistic electrons cannot explain the observed $\gamma$-ray
emission for two different reasons. First, the strong magnetic field produced by
the CR-induced streaming instability forces the number density of relativistic
electrons to be too small to explain the $\gamma$-ray emission as due to ICS on
the ambient photons. Second, even if we arbitrarily reduced the magnetic field
strength, enhancing at the same time the electron number density in order to fit
the TeV $\gamma$-rays with ICS emission, we could not account for the GeV
$\gamma$-rays because both the spectral slope and the flux would be incompatible
with the Fermi-LAT data.
Also non-thermal electron bremsstrahlung has to be ruled out, because it
provides a flux two order of magnitudes lower than the Fermi-LAT detection, and
cannot be arbitrarily enhanced without over-predicting both 
the TeV and the X-ray emission. 
Kepler, due to its larger distance, it is fainter than Tycho and therefore not
detected in the $\gamma$-ray band by the current generation of telescopes.
However, the predicted TeV flux should be observable by the Cerenkov Telescope
Array.

\begin{thebibliography}{}

\bibitem[\protect\citeauthoryear{Drury et al.}{1994}]{dav94} 
        Drury, L.~O'C, Aharonian, F., V\"{o}lk, H.~J.,
        {\it A$\&$A} {\bf 287}, 959- (1994)

\bibitem[\protect\citeauthoryear{Ellison et al.}{2007}]{Ellison07}
        Ellison, D.~C., Patnaude, D.~J., Slane, P., Blasi, P., Gabici, S.,
        {\it ApJ} {\bf 661}, 879- (2007)

\bibitem[\protect\citeauthoryear{Giordano et al.}{2011}]{Giordano11} 
        Giordano, F. et al., {\it ApJ} {\bf 744}, 2- (2012)

\bibitem[\protect\citeauthoryear{Acciari et al.}{2011}]{Acciari11} 
        Acciari, V.~A. et al., for the VERITAS collaboration, 
        {\it ApJ} {\bf 730}, L20- (2011)

\bibitem[\protect\citeauthoryear{Morlino \& Caprioli}{2011}]{morlino11} 
        Morlino, G. \& Carpioli, D., {\it A\&A} {\bf 538}, 81- (2012)

\bibitem[\protect\citeauthoryear{Krause et al.}{2008}]{Krause08} 
        Krause, O. et al., {\it Nature} {\bf 456}, 617- (2008)

\bibitem[\protect\citeauthoryear{Reynolds et al.}{2007}]{Reynolds07} 
        Reynolds, S. P., Borkowski, K. J., Hwang, U., et al., 
        {\it ApJ}  {\bf 668}, L135- (2007)

\bibitem[\protect\citeauthoryear{Truelove \& Mc Kee}{1999}]{TMK99}
        Truelove, J.~K. and Mc Kee, C.~F., {\it Apj~Supplement Series} 
        {\bf 120}, 299- (1999)

\bibitem[\protect\citeauthoryear{Caprioli et al.}{2010b}]{dsax0}
        Caprioli, D., Amato, E., P. Blasi, {\it APh} {\bf 33}, 307- (2010)


\bibitem[\protect\citeauthoryear{Caprioli et al.}{2009}]{jumpkin}
        Caprioli, D., {\it JCAP} {\bf 7}, 38- (2012)

\bibitem[\protect\citeauthoryear{Reynolds \& Ellison}{1992}]{ReyEllison92} 
        Reynolds, S.~P., \& Ellison, D.~C., {\it ApJ} {\bf 339}, L75- (1992)

\bibitem[\protect\citeauthoryear{Allen et al}{1999}]{allen99}
        Allen, G.~E., Gotthelf, E.~V., Petre, R., {\it Proceedings of the 26th
        ICRC} {\bf 3}, 480- (1999) 

\bibitem[\protect\citeauthoryear{Cassam-Chena\"i et al.}{2007}]{Gamil07} 
        Cassam-Chena\"i, G., Hughes, J.~P., Ballet, J., Decourchelle, A.,
        {\it ApJ} {\bf 665}, 315- (2007)

\bibitem[\protect\citeauthoryear{Vink}{2008}]{Vink08} 
        Vink, J., {\it ApJ} {\bf 689}, 231- (2008)







\end{thebibliography}
\end{document}